\documentclass[preprint]{aastex}

\usepackage{graphicx,epsfig,subfigure}
\usepackage{amssymb,amsmath}
\usepackage{epstopdf}

\shorttitle{A framework for interpreting fast radio transients search experiments: application to the V-FASTR experiment}
\shortauthors{C. M. Trott}

\begin{document}

\title{A framework for interpreting fast radio transients search experiments: application to the V-FASTR experiment}


\author{Cathryn M. Trott\altaffilmark{1}, Steven J. Tingay, Randall B. Wayth}
\affil{International Centre for Radio Astronomy Research, Curtin University, Bentley WA 6845, Australia}
\email{cathryn.trott@curtin.edu.au}
\author{David R. Thompson}
\affil{Jet Propulsion Laboratory, California Institute of Technology, 4800 Oak Grove Drive, Pasadena, CA 91109, USA}
\author{Adam T. Deller}
\affil{ASTRON, Oude Hoogeveensedijk 4, 7991 PD Dwingeloo, The Netherlands}
\author{Walter F. Brisken}
\affil{NRAO, PO Box 0, Socorro, NM 87801, USA}
\author{Kiri L. Wagstaff, Walid A. Majid, Sarah Burke-Spolaor}
\affil{Jet Propulsion Laboratory, California Institute of Technology, 4800 Oak Grove Drive, Pasadena, CA 91109, USA}
\author{Jean-Pierre R. Macquart, Divya Palaniswamy}
\affil{International Centre for Radio Astronomy Research, Curtin University, Bentley WA 6845, Australia}

\altaffiltext{1}{Centre of Excellence for All-Sky Astrophysics (CAASTRO)}


\begin{abstract}
We define a framework for determining constraints on the detection rate of fast transient events from a population of underlying sources, with a view to incorporating beam shape, frequency effects, scattering effects, and detection efficiency into the metric. We then demonstrate a method for combining independent datasets into a single event rate constraint diagram, using a probabilistic approach to the limits on parameter space. We apply this new framework to present the latest results from the V-FASTR experiment, a commensal fast transients search using the Very Long Baseline Array (VLBA). In the 20~cm band, V-FASTR now has the ability to probe the regions of parameter space of importance for the observed Lorimer and Keane fast radio transient candidates, by combining the information from observations with differing bandwidths, and properly accounting for the source dispersion measure, VLBA antenna beam shape, experiment time sampling, and stochastic nature of events. We then apply the framework to combine the results of the V-FASTR and ATA Fly's Eye experiments, demonstrating their complementarity. Expectations for fast transients experiments for the SKA Phase I dish array are then computed, and the impact of large differential bandwidths is discussed.
\end{abstract}



\section{Introduction}
Probing the time domain is emerging as one of the exciting goals of current and future radio instruments. Detection and characterization of \textit{fast} transient events (those varying on time scales shorter than the correlator time scale in a typical imaging pipeline, and usually sub-second) is expected to probe some of the most energetic and dynamic astrophysical events, particularly for extragalactic sources. Pulsars are the most observed rapidly-varying radio source, where high time resolution voltage capture, and folding of multiple pulse profiles, are used to form an incoherent or coherent detection. In addition to the pulsar source population, which is well-studied, there are hopes for detection and characterization of other astrophysical populations, including other emission mechanisms from neutron stars \citep[e.g., magnetars, RRATs, ][]{keane10}, sporadic emission from flare stars and planets, radio emission from gamma ray bursts (GRBs), and coherent processes involving AGN \citep{macquart10, cordes03}. Within the radio spectrum, the 20~cm waveband has been best studied, with two interesting candidate extragalactic sources observed with the multibeam receiver of the Parkes radiotelescope \citep{lorimer07,keane11}. These first results, and the source localization limitations of single-element systems, have prompted keen interest in this field.

There have been a variety of fast transients experiments undertaken in the past thirty years, with a concentration of resources in the past decade \citep{deneva09,siemion12,keane10,keane11,burke11}, and plans for future experiments \citep{lonsdale09,stappers11,bhat11,macquart10}. These experiments have operated in a range of modes, yielding constraints probing different parts of the astrophysical parameter space, as well as providing alternative strategies for detection. The modes include different instrument types (single-element and multiple-element), experimental set-up (frequency, bandwidth, channel width, temporal resolution), and different signal capture strategies (incoherent addition of power from each element, coherent addition of voltages, fly's eye pointing of elements to different fields). These modes aim to balance field-of-view with sensitivity, and allow optimal radio frequency interference (RFI) mitigation techniques for each experiment.

V-FASTR, The VLBA Fast Radio Transients Experiment \citep{wayth11,thompson11,wayth12}, is a commensal fast transient experiment, using the ten 25~m dish antennas of the North American Very Long Baseline Array (VLBA) network \citep{napier94}. A full experiment and system description is available in these publications, and we present a brief description here. V-FASTR operates continuously, searching for fast transients in real time commensally with regular VLBA data acquisition. Since fields are selected based on the science goals of high angular resolution astronomy rather than transient science, V-FASTR dataset consists of observations of a range of targets with different receiver setups and system bandwidths. The resulting selection effects and biases are discussed by Wayth et al. (2012). The ten VLBA antennas are separated by a maximum baseline of $\sim$8000~km, offering exceptionally high angular resolution when used in cross-correlation mode. In contrast, V-FASTR incoherently adds the auto-correlated power from each antenna. The field of view is the same as that of an individual element, with sensitivity scaling as $\sqrt{N_{ant}}S$, where $S$ is the sensitivity for an individual element and $N_{ant}$ is the number of available antennas.

The distributed nature of the array means that interfering sources are not correlated between antennas, enabling excellent RFI rejection. The V-FASTR pipeline optimizes the removal of RFI using custom signal processing techniques \citep{thompson11}. \citet{wayth12} present results for a subset of V-FASTR data (only data with 64~MHz bandwidth collected before 2012 May). Here, we update their results for a wider range of bandwidths and more recent data. We also develop a framework that allows improved fast transient event rate constraints, by combining observations with different bandwidths.

Apart from phenomena arising from neutron stars (pulsars and RRATs), most fast transients experiments have been unable to detect populations of sources. Typically, after completion of an experiment, an event rate constraint plot is generated, which seeks to outline the region of flux density versus event rate density that has been excluded by the absence of astrophysical detections. The transient event space is inherently multi-dimensional; in addition to source and propagation parameters (e.g., luminosity, distance, dispersion, time scale, sky temperature spatial fluctuations), there are the effects of the measurement system, which include instrument parameters (field-of-view, sensitivity, bandwidth, frequency), and detection parameters (incoherent versus coherent, single versus multiple element, sampling time scale, dispersion measure trials). The latter can have a significant impact on the ability to detect and confirm events, particularly with reference to excising radio frequency interference (RFI).

Up to this point, there has been no clear framework for incorporating data from different experiments. As part of this, there is no existing framework for (1) detailed study of the impact of varying instrument sensitivity (frequency-dependent noise, beam shape), (2) incorporating the effects of experimental parameters (frequency differences, bandwidth, temporal sampling), or (3) incorporating detection performance degradation from the detection strategy (scatter broadening, boxcar templates, RFI excision choices). In this paper we develop a framework to account for differences in experimental and source parameters, allowing a scaling of the results of an experiment to a common standardized quantity. Having scaled two experiments, we then demonstrate how to combine the results into a single constraint plot, using a probabilistic description of the ability of an experiment to include or exclude regions of parameter space. This framework is applied to the V-FASTR data accumulated to 2012, October 30, allowing the combination of observations with different bandwidths, and correctly incorporating the frequency-dependent beam shape of the VLBA antennas. We then demonstrate how to combine these data with the results of other, similar, surveys, yielding an event rate constraint curve in the 20~cm band with combined V-FASTR and Allen Telescope Array (ATA) Fly's Eye results \citep{siemion12}. Finally we compute the expected constraint diagram for the specified SKA Phase I dish system, and 461 hours of observation \citep{dewdney10}.

\section{Impact of observation parameters on event rate constraints}
\subsection{Bandwidth and beam shape}
We begin by defining the signal power and noise power in a time series, which has been de-dispersed at the correct dispersion measure. We assume the total power is the sum of the power in each channel, normalized by the number of channels (average power). We will initially assume that the source is located at the beam centre, and there is no frequency dependence to the signal or noise samples:
\begin{eqnarray}
P_S = \frac{1}{N_{\rm ch}}\displaystyle\sum_{i=1}^{N_{\rm ch}} P_i\\
P_N = \frac{S_i}{\sqrt{N_{\rm ch}}},
\end{eqnarray}
where the sum extends over the $N_{\rm ch}$ spectral channels, and $S_i$ is the noise uncertainty in each power sample, in each channel. For the incoherent mean of power from multiple antennas, the signal is unchanged, and the noise reduced.
Incorporating frequency-dependent signal and noise, extending to a continuous frequency coverage, and incorporating a frequency-dependent beam, $B(\nu)$ yields:
\begin{eqnarray}
P_S = \frac{1}{N_{\rm ch}N_{\rm ant}\Delta\nu} \displaystyle\sum_{j=1}^{N_{\rm ant}} \displaystyle\int_{BW} P(\nu)B_j(\nu)d\nu\\
P_N = \frac{1}{N_{\rm ch}N_{\rm ant}\sqrt{\Delta\nu}}\sqrt{\displaystyle\sum_{j=1}^{N_{\rm ant}} \displaystyle\int_{BW}S_{j,{\rm sys}}^2(\nu)d\nu},
\end{eqnarray}
where the integral extends over bandwidth, $BW$, the system noise, $S_{j,{\rm sys}}(\nu)$ is the noise in a frequency channel at frequency, $\nu$, for antenna $j$, and $N_{\rm ant}$ and $\Delta\nu$ are the number of antennas and channel spectral resolution, respectively. We have allowed for the possibility of non-identical antennas. Herein we assume identical antennas, with the understanding that the full expression can be used in the general case.

For a candidate to be considered a detection, the signal power needs to exceed the threshold, given by the noise power multiplied by some signal-to-noise ratio value, $C$. A signal will be detected if it meets the following criterion:
\begin{equation}
\frac{\sqrt{N_{\rm ant}} \displaystyle\int_{BW} P(\nu)B(\nu)d\nu}{\Delta\nu} > C \frac{\sqrt{\displaystyle\int_{BW}S_{\rm sys}^2(\nu)d\nu}}{\sqrt{\Delta\nu}}.
\label{signal_noise_eqn}
\end{equation}
We represent the strength of the underlying astrophysical signal as a power-law in frequency, yielding $P(\nu) = S_0(\nu/\nu_0)^\alpha$, where $S_0$ is the flux density at reference frequency, $\nu_0$. The inequality in equation (\ref{signal_noise_eqn}) becomes an equality at the minimum source flux (at reference frequency $\nu_0$), $S_{min} = S_0$. The foregoing formalism is straightforwardly generalized to include the angular dependence of the beam shape. The noise power is unchanged across the beam, but the signal power is attenuated by the beam response. The minimum detectable flux density at angle $\theta$ from the beam centre, $S_{min}(\theta)$ is given by:
\begin{equation}
S_{min}(\theta) = \frac{C \sqrt{\Delta\nu \displaystyle\int_{BW}S_{\rm sys}^2(\nu)d\nu}}{\sqrt{N_{\rm ant}} \displaystyle\int_{BW} \left(\frac{\nu}{\nu_0}\right)^\alpha B(\nu,\theta)d\nu}.
\end{equation}
As expected, larger bandwidths and smaller thresholds yielder lower minimum detectable source fluxes.

It is common to represent a homogeneous population of events by a single point in the 2D Cartesian plane, with the vertical axis representing signal magnitude and the horizontal representing rate per area per time.  A survey that detects a single event provides an observed flux density and rate.  With enough observations one could confidently estimate both the expected flux density and rate. On the other hand, a survey that does not observe any events provides some information to exclude parts of the event rate plane.  For convenience, it is common to plot an exclusion boundary for a "null result" sky survey by the isocontour of parameter combinations that would have produced a single event in expectation, thereby defining a locus of Cartesian points with co-ordinates:
\begin{equation}
[{\textsf{ Rate per area per time}}, S_{min}(\theta)] = \left[\frac{1}{\displaystyle\sum_{S_i<S_{min}(\theta)} 2\pi\theta_i\Delta\theta\Delta{t}},S_{min}(\theta) \right],
\end{equation}
where the area of an annulus of width $\Delta\theta$ has been incorporated.

\subsubsection{Scatter broadening losses}
Due to multi-path propagation through the ISM, pulses are broadened in time, with characteristic timescales that are heavily frequency-dependent \citep[$\nu^{-4}-\nu^{-4.4}$,][]{cordes03}. Pulse broadening reduces the instantaneous signal strength, as the pulse energy is distributed over a longer timescale, leading to a degradation in the achievable signal-to-noise ratio, compared with the intrinsic pulse. For a square pulse of intrinsic width, $W$, convolved with an exponential decay with characteristic timescale, $\tau$, the loss in SNR is given by:
\begin{equation}
\frac{SNR_{\tau}}{SNR_{\rm optimal}} = \sqrt{1-\beta+\beta\exp{(-1/\beta)}} \equiv \epsilon ,
\end{equation}
where $\beta = \tau/W$\footnote{Obtained by integration of a square pulse convolved with an exponential tail, assuming a matched filter (optimal) detection template. The detection SNR for the matched filter and a general pulse shape, is given by:
\begin{equation}
d = \frac{\sqrt{\int_W s(t)^2dt}}{\sigma},
\label{snr_mf}
\end{equation}
where $s(t)$ is described by a square pulse convolved with a decaying exponential scatter broadening function.}. The scattering timescale in the Galaxy, $\tau=\tau({\rm DM},\nu)$, is given empirically by \citep{cordes03}:
\begin{equation}
\log{\tau} = -3.72 + 0.411 \log{\rm DM} + 0.937(\log{\rm DM})^2 - 4.4\log{\nu_{\rm GHz}} \hspace{2mm} \mu{\rm s},
\label{tau_d}
\end{equation}
with substantial empirical scatter.
The magnitude of the effect is dependent on the dispersion measure of the signal, DM, and the frequency, $\nu$. 
The degradation factor, $\epsilon$, is an approximation due to the scatter in the empirical relationship between DM, frequency and temporal broadening, and the limited availability of evidence for extragalactic sources.

Appropriate application of this scaling relation requires a DM value to be assumed. One can incorporate this detection performance degradation into the event rate constraint plot. This scaling relation also requires knowledge of the intrinsic pulse width, $W$. The sampling time scale of the experiment, $\Delta{t_s}$, is a useful proxy for this quantity (for short pulses, the sampling time scale is the crucial scaling in the system).

\subsection{Intrinsic pulse width and temporal sampling}
An experiment that uses a temporal sampling time that differs from the intrinsic pulse width could suffer a loss of detection performance, due to the spreading of signal over time. If the sampling time scale is longer than the intrinsic pulse width (or, the observed pulse width after propagation through the plasma medium), the signal strength is degraded (averaged), and the pulse is less detectable. If the sampling time scale is shorter than the pulse width, detection performance benefits from averaging over multiple timesteps to capture all of the pulse energy (e.g., boxcar averaging performed in V-FASTR, and other experiments).

As a proxy for the signal contained in an event, we can form a composite quantity,
\begin{equation}
S_{min} \sqrt{\Delta{t_s}} \approx S_{\rm actual}\sqrt{W} \hspace{0.5cm} {\rm Jy.s^{1/2}},
\end{equation}
where $\Delta{t_s}$ is the experiment sampling time scale, and $W$ is the intrinsic pulse width. This expression relates the true, unknown pulse width and actual source flux, to the sampling time scale and the measured flux, and provides a quantity that accounts for a range of intrinsic pulse widths (e.g., plotting the experimental minimum flux density and timescale provides an approximate measure of the true flux density and pulse width). Note that this expression differs from the oft-used energy-like quantity, $S\Delta{t}$, by a square-root dependence on time. In this work, we are interested in determining the impact on signal-to-noise ratio of different effects, with a view to applying this framework to experiments with SNR thresholds that are used to define detections. The SNR scales as the square-root of the temporal sampling.


\section{A new combined metric}
We have formed scaling relations for incorporating a number of basic properties into an understanding of our event rate constraints. We can now combine these to form a quantity that provides a measure of the true signal at reference frequency, $\nu_0$ (within the band), and beam angle, $\theta$, for a given combination of experimental parameters ($BW$, $C$, $\Delta{t_s}$, $\Delta{t}$) and source parameters ($\alpha$, DM):
\begin{eqnarray}\label{all_eqn}
S_{\rm actual}(\nu_0,\theta;BW,C,\alpha,\Delta{t_s},{\rm DM},\Delta{t})\sqrt{W} =
\frac{C \sqrt{\Delta{t_s} \Delta\nu \displaystyle\int_{BW}S_{\rm sys}^2(\nu)d\nu}}{\epsilon \sqrt{N_{\rm ant}} \displaystyle\int_{BW} \left(\frac{\nu}{\nu_0}\right)^\alpha B(\nu,\theta) d\nu}.
\end{eqnarray}
By varying the DM, this yields a family of curves in the event rate diagram. A more general metric incorporates the detection performance degradation for detecting a high DM signal, and is given by:
\begin{equation}
\epsilon S_{{\rm actual}}(\nu_0,\theta)\sqrt{W} = \sqrt{\Delta{t_s}}\frac{C \sqrt{\Delta\nu \displaystyle\int_{BW}S_{\rm sys}^2(\nu)d\nu}}{\sqrt{N_{\rm ant}} \displaystyle\int_{BW} \left(\frac{\nu}{\nu_0}\right)^\alpha B(\nu,\theta) d\nu},
\label{final_equation}
\end{equation}

As before, variation of the beam angle, $\theta$, yields a locus of points in the event rate plane with corresponding sky areas:
\begin{align}
[{\textsf{ Rate per area per time}}, \epsilon S_{min, {\rm actual}}(\nu_0,\theta)\sqrt{W}] \\\nonumber
&= \left[\frac{1}{\displaystyle\sum_{S_i<S_{min}(\theta)}2\pi\theta_i\Delta\theta\Delta{t}},\epsilon S_{min, {\rm actual}}(\nu_0,\theta)\sqrt{W} \right].
\end{align}
The event rate constraint curve plots the number of events per sky area per unit time against the minimum detectable flux density at frequency $\nu_0$ times the square-root sampling time scale (as a proxy quantity for the actual flux density, times the square-root of the intrinsic pulse width, times the performance degradation due to temporal smearing of the signal).

\section{Combining datasets}\label{all_together}
Having demonstrated a scaling relation to take the parameters of any experiment and scale them to a common quantity (equation \ref{all_eqn}), we can now explore how to compare and combine independent results into a single constraint curve. In the simplest case, where the minimum detectable flux density for two experiments is the same, and no transients have been found, one can simply add the FOV$\times$time quantities (survey volume) for each survey to find the total survey volume to be used in the abscissa of the event rate plot. However, if two surveys have different detection thresholds, it is unclear how the information can be combined.

This issue highlights a broader problem in the formation and interpretation of event rate constraint plots: the constraints are inherently probabilistic, and one cannot assign absolute boundaries beyond which a survey has excluded a region of parameter space. Instead, the noise in the dataset, and the presumably Poisson nature of the occurrence of fast transients, mean that the detection of an event is a single realization of an underlying statistical distribution.

We begin by developing the probability distribution functions for the two plotted quantities in the event rate diagram: source flux density and number of events. We treat these functions separately. We aim to understand the chance of the null hypothesis: that there are no real events. Rejecting the null hypothesis at some level of significance indicates the probabitity that finite sensitivity and a threshold, or an unlucky timing of the experiment has caused us to not observe anything.
\newline\noindent
{\bf Source flux density}\\
\noindent
A source with true flux density, $S$, is embedded within statistical noise. Hence, the measured value of the flux density is a realization of a Gaussian-distributed random variable, with mean value, $\mu=S$, and variance given by the dataset noise, $\sigma^2$. The probability a source of true flux density $S$ will be detected above threshold, $C\sigma$, is given by the cumulative distribution function (CDF):
\begin{eqnarray}
P(X>C\sigma) &=& \displaystyle\int_{C\sigma}^{\infty} \mathcal{N}(S,\sigma^2) dx\\
&=& \frac{1}{2} + \frac{1}{2}{\rm erf}{\left(\frac{S-C\sigma}{\sqrt{2}\sigma} \right)}
\label{gauss_eqn}
\end{eqnarray}
where erf is the error function, and $\mathcal{N}(\mu,\sigma^2)$ denotes a Gaussian distribution with mean, $\mu$, and variance, $\sigma^2$. The cumulative probability in equation \ref{gauss_eqn} describes the exclusion regions. The cumulative probability that an event is not detected (due to noise) is the complementary function, $1-P(X>C\sigma)$.
\newline\noindent
{\bf Expected event rate}\\
\noindent
Real astrophysical events are assumed to be randomly distributed on the sky, with a given (unknown) mean frequency of occurrence, but random actual timing. This distribution is governed by the law of rare events, and follows the Poisson distribution (observation of zero fast transient events during an experiment does not imply that none would have been observed in an identical experiment). The probability of observing $k$ events, given a mean (expected) number, $\lambda$, is given by:
\begin{eqnarray}
P(k;\lambda) &=& \frac{k^\lambda\exp{-\lambda}}{k!}\\\label{poisson_eqn}
&=& \exp{-\lambda} \hspace{0.2cm} |_{k=0},
\end{eqnarray}
implying that for $k=0$ (zero observed events), there is a non-zero probability of seeing one or more events in an identical experiment. In the event rate diagram, the Poisson probability distribution is plotted at,
\begin{equation}
X = \lambda/(FOV \times \Delta{t}),
\end{equation}
where $X$ is the \textit{expected} number density of events (compared with the \textit{observed} number, typically plotted in an event rate diagram). If we have observed zero events, the chance that we should have observed at least one is given by the complementary function, $1-Pr(k=0;\lambda)$.

\subsection{Allowed region of parameter space}
\begin{figure}
\begin{center}
\includegraphics[scale=0.9]{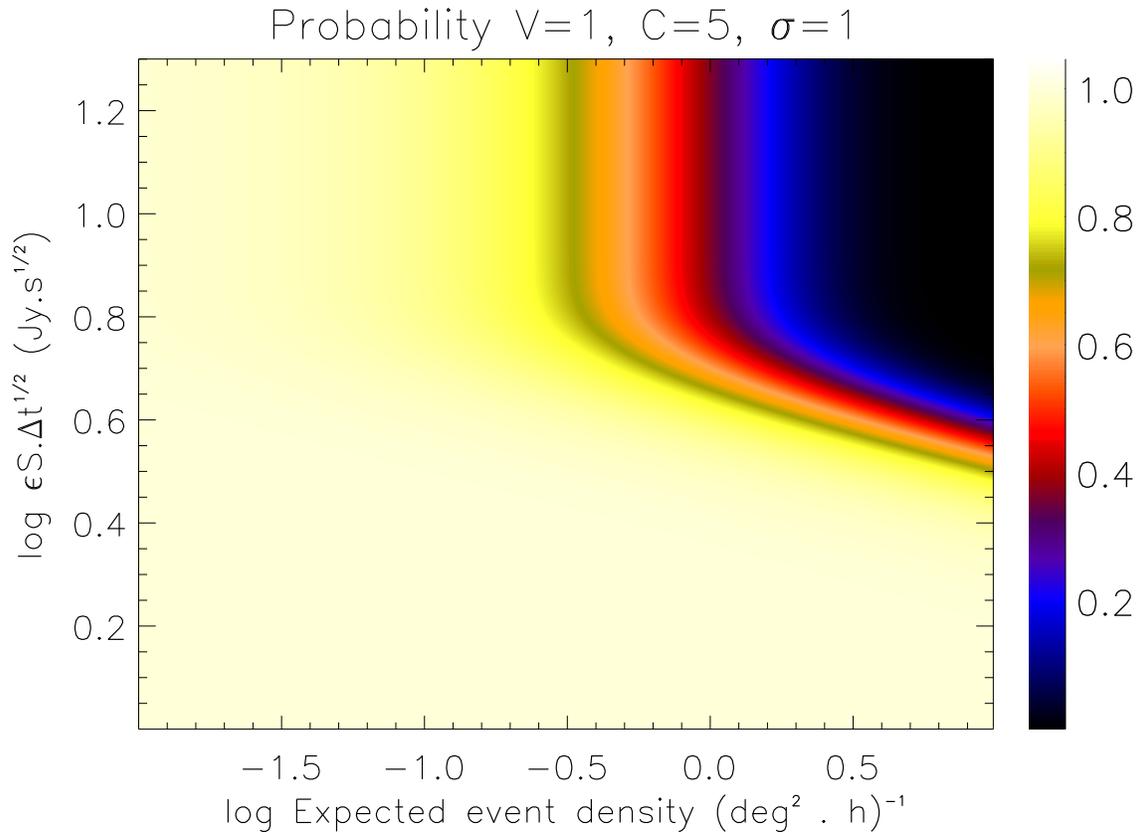}
\caption{Example probability event rate plot, displaying the probability that events are present, but not detected (due to noise or the stochastic nature of event timing). Threshold is set to $C$=5 with $\sigma=1$, $\Delta{t_s}=1$~s, FOV.$\Delta{t}$=1 deg.$^2$s.}\label{new_event_rate_fig}
\end{center}
\end{figure}

There are two possibilities when interpreting a null result experiment: (1) that no events were observed because there was nothing there (the null hypothesis), or (2) that fast transients are present, but noise or stochastic timing prevented observation. These two hypotheses have a total probability of unity. The probability of the null hypothesis, $P(H_0)$, corresponds to the chance that at least one source would have been detected, and is given by\footnote{Mathematically, this expression sums over possible non-zero true events, and multiplies the probability of $n$ occurring by the probability that at least one would be above the threshold.}:
\begin{eqnarray}
P(H_0) &=& \textsf{at least one signal is above the threshold when $n$ events do occur}\\
&=& \displaystyle\sum_{n>1} (1 - P(X<C\sigma)^n) \times (P(k=n;\lambda)),
\end{eqnarray}
and the probability that events do occur, but were not detectable, is given by:
\begin{equation}
1 - P(H_0) = 1 - \displaystyle\sum_{n>1} (1 - P(X<C\sigma)^n) (P(k=n;\lambda)).
\label{prob_eqn}
\end{equation}
Figure \ref{new_event_rate_fig} demonstrates the new event rate plot. The quantity plotted is the probability that an event in this part of parameter space would not be detected by our experiment (hence, low probabilities correspond to traditional areas of exclusion --- frequent, bright events).
At the beam centre, where the signal is not attenuated, the experiment is able to exclude regions at lower flux densities, but at the expense of small FOV (larger expected event rate required).

The combined probability for multiple experiments considers the joint probability that events occur but are not observed in either experiment.
The total probability for $N$ \textit{independent} experiments is given by:
\begin{equation}
P_{\rm Tot} = \displaystyle\prod_{i=1}^N (1-P_i(H_0)).
\label{combined_prob_eqn}
\end{equation}
This expression treats the experiments independently\footnote{Independence is appropriate for the case of combining information from different pointings, or different instruments. For constructing an event rate plot for the same pointing, where different areas of the beam have different minimum detectable flux densities, only the new part of the field-of-view is independent (the incremental information added by a small area is independent of the information available from the existing area use in the plot). To construct an event rate constraint diagram across the beam, we treat each area annulus independently, and incrementally combine information.}, and creates more stringent event rate constraints as more information is added. For a weak experiment (high minimum flux density, for example), the null hypothesis has zero probability for a large amount of parameter space, yielding little additional constraint when combined with a stronger experiment. Note that this expression \textit{may not} give the same answer as taking two non-identical experiments and summing the total survey volume for a stronger constraint. This is because the probability description includes the sensitivity of an experiment, which, in general, is not the same for two experiments.

We now have all the tools to be able to combine constraints from different experiments. It remains to discuss the types of experiments that are meaningfully combined, and those that are not. In general, any experiment could be combined with any other, and all of the differences incorporated into an overall framework. However, in practise our new framework only considers noise level, beam shape and signal differences. Results from the same instrument and different bandwidths can be meaningfully combined, but this framework may be too simplistic to combine single element with multiple element experiments. This is because it considers the data to be clean of RFI, justifying the use of Gaussian statistics to describe the data. In practice RFI can hurt an investigation in one of two ways: first, by causing false positive detections, which does not affect the sensitivity limit (though it may raise the burden of candidates to examine); and second, by artificially raising the background signal used to compute $C\sigma$, and as a result reducing the actual sensitivity. In general, RFI excision with single element experiments is difficult, due to a lack of independent information, and datasets typically contain some level of RFI. Indeed, even V-FASTR is subject to RFI, and occasionally RFI spikes add incoherently to form a detection. The different strategies employed by these experiments lead to differing RFI statistics in the data. Because our framework does not consider these aspects, it cannot provide a level playing field to combine all experiments.

This highlights a more general discussion about the performance of a detector. In this work, we have focussed our attention on forming a more realistic event rate constraint diagram, with no reference to the design and performance of a given detection strategy (we have simply taken a threshold signal-to-noise ratio). In practise, the threshold, $C$, is chosen to balance false positive detections with false negatives (true astrophysical transients are not detected). The overall performance of a detection strategy can be quantified by exploring this balance as the threshold is varied. \citet{thompson11} explore the use of receiver operator characteristic curves (plotting false positive fraction versus false negative fraction) to quantify the performance of different detection strategies. In the V-FASTR context, antennas are combined in different ways to discriminate true signals from RFI. The data are inherently non-Gaussian, due to the underlying RFI environment, and \citet{thompson11} incorporate this information into the data statistics to form useful detectors. In a similar way, the assumption of Gaussian-distributed noise used here (equation \ref{gauss_eqn}) could be extended to use any general distribution function (including the statistics of remaining RFI), and the event rate constraint curve altered to reflect this\footnote{V-FASTR does have subtle non-Gaussian effects due to excising large-magnitude datapoints during RFI-contaminated observations. However, there is not a well-defined analytical way to model the resulting sensitivity}. With a good understanding of the RFI environment of each experiment, a broader class of experimental results could be combined meaningfully. This generalization is beyond the scope of this work.


We use the description in equation \ref{all_eqn} to standardize each experiment, taking into account performance losses due to scattering, bandwidth and sampling time scale. We then form the probabilistic description of the event rate constraint, according to equation \ref{prob_eqn}, and combine according to equation \ref{combined_prob_eqn}. Note that this same formalism is used to combine results from different beam angles, for the same experiment. The validity of this assumption rests with the assumption that all look directions are equivalent. 

\section{Application: V-FASTR constraints}
The V-FASTR experiment has produced event rate constraints for ten observing wavebands, ranging from central frequencies of 325~MHz to 86.2~GHz. Over this frequency range, the beam size, SEFD, and time-on-sky, vary considerably. We will first explore how the previously-used event rate plot is modified by the inclusion of experimental parameters, and then produce the new plots for four interesting wavebands. As with previous estimates, we assume that the antennas are identical. For simplicity, we will also assume that there is no variation of system temperature over the bandwidth of an observation (although we have developed the framework to account for this).

Table \ref{results_table} shows the total time-on-sky results, by receiver and bandwidth, for V-FASTR, as of 2012 October 30. The total time is divided into three bandwidths, $BW$ = [64~MHz, 128~MHz, 256~MHz].
\begin{table}
\begin{minipage}{15cm}
\begin{tabular}{|c|c|ccc|c|c|}
\hline Receiver & $\Delta{t}_{\rm Tot}$ (h) & $\Delta{t}_{\rm 64}$ & $\Delta{t}_{\rm 128}$ & $\Delta{t}_{\rm 256}$ & ${N_{\rm ant}}$ & SEFD (Jy)\footnote{Obtained from http://www.vlba.nrao.edu/astro/obstatus/current/node7.html} \\ 
\hline 90cm &	8.33 &	8.27 &	0.06 &	0.00 &	9.52 & 2227\\
50cm &	0.06 &	0.00 &	0.06 &	0.00 &	10.00 & 2216\\
20cm &	704.57 &	539.87 &	43.53 &	121.17 &	9.44 & 296\\
13cm &	197.38 &	188.52 &	4.27 &	4.60 &	9.03 & 322\\
6cm &	48.11 &	27.99 &	9.20 &	10.92 &	9.65 & 210\\
4cm &	732.99 &	423.73 &	215.68 &	93.57 &	10.13\footnote{At 4~cm, additional antennas are sometimes available, when the VLBA is used for geodetic purposes.} & 307\\
2cm &	385.47 &	287.36 &	6.09 &	92.02 &	9.57 & 550\\
1cm &	480.57 &	411.98 &	43.27 &	25.31 &	9.83 & 502\\
7mm &	328.01 &	253.86 &	2.41 &	71.74 & 9.45 & 1436\\
3mm &	29.91 &	17.92 &	0.00 &	11.99 &	7.72 & 4000\\ 
\hline 
\end{tabular}
\caption{V-FASTR time-on-sky results for each receiver, as of October 30, 2012. The total time is split into time with bandwidths, $BW$ = [64~MHz, 128~MHz, 256~MHz]. Also shown is the average number of antennas available with that receiver, $N_{\rm ant}$, and the SEFD used in the noise calculations.}\label{results_table}
\end{minipage}
\end{table}
Some frequencies have substantial time at the larger bandwidth. We can use the framework developed here to combine the information from different frequencies.

The VLBA antennas are 25~m dishes, with beam patterns that are well-described by Airy disks \citep{dhawan02}:
\begin{equation}
B(\nu,\theta) \propto \left(\frac{J_1(u)}{u}\right)^2 = \left(\frac{cJ_1(\pi{D}\nu\sin{\theta}/c)}{\pi{D}\nu\sin{\theta}}\right)^2,
\end{equation}
where $J_1(u)$ is the Bessel function of the first kind. The Airy function passes through multiple nulls off-axis, and a source located close to a null will have a corresponding large minimum detectable flux density\footnote{The beam pattern of a VLBA antenna is well approximated by the Airy function, which possesses a series of nulls which may be determined approximately from the asymptotic form of the Bessel function, $J_m(z)$ for $z \gg | m^2 - 1/4 |$:
\begin{eqnarray}
J_m (z) \approx \sqrt{\frac{2}{\pi z}} \cos \left(z - \frac{m \pi}{4} - \frac{\pi}{4} \right). 
\end{eqnarray} 
Thus the Airy function possesses a null at the points $x = n \pi$, where $n=1,2,3, \ldots$.  (There is no null at $n=0$ since $\lim_{x \rightarrow 0} (2 J_1(x)/x)^2 = 1$.) It is evident that the spectrum of a source that is observed at an angular position $\theta_0$ well off axis (i.e. $\theta_0 > 1/k a$) and over a large bandwidth will be subject to a large number of oscillations, due to the oscillations in the beam response over the wide range of $k$ (wavenumber).  We can quantify this by examining the number of nulls imprinted in the observed source spectrum due to oscillations in $B(\theta)$.  Since a null occurs every time $k a \theta_0$ changes by $\pi$, the number of nulls for an observation between frequencies $\nu_1$ and $\nu_2$ is 
\begin{eqnarray}
N = \left\lfloor \frac{k_2 a \theta_0 - k_1 a \theta_0}{\pi} \right\rfloor = \left\lfloor \frac{2 \theta_0 a}{c} (\nu_2 - \nu_1) \right\rfloor.
\end{eqnarray}
Note that $N$ depends only on the total bandwidth, and not on the relative bandwidth $\Delta \nu/\nu$.  
Thus the number of oscillations present in the detection spectrum indicates the offset of the source from the pointing centre. However, for a fixed $\theta_0$ obviously the beam width is smaller at high frequency, and so for a source well off axis the overall amplitude of $B(\theta)$ will be much smaller at high frequencies, rendering the source harder to detect, for a given (constant) flux density.}.
We truncate the VLBA beam beyond the third Airy null ($u\sim$ 13.3), and set the gain to a constant value to the horizon, following the work of \citet{deneva09} and \citet{wayth12}.

\begin{figure}
\begin{center}
\subfigure[Variation with source DM (BW = 64~MHz).]{\includegraphics[scale=0.75]{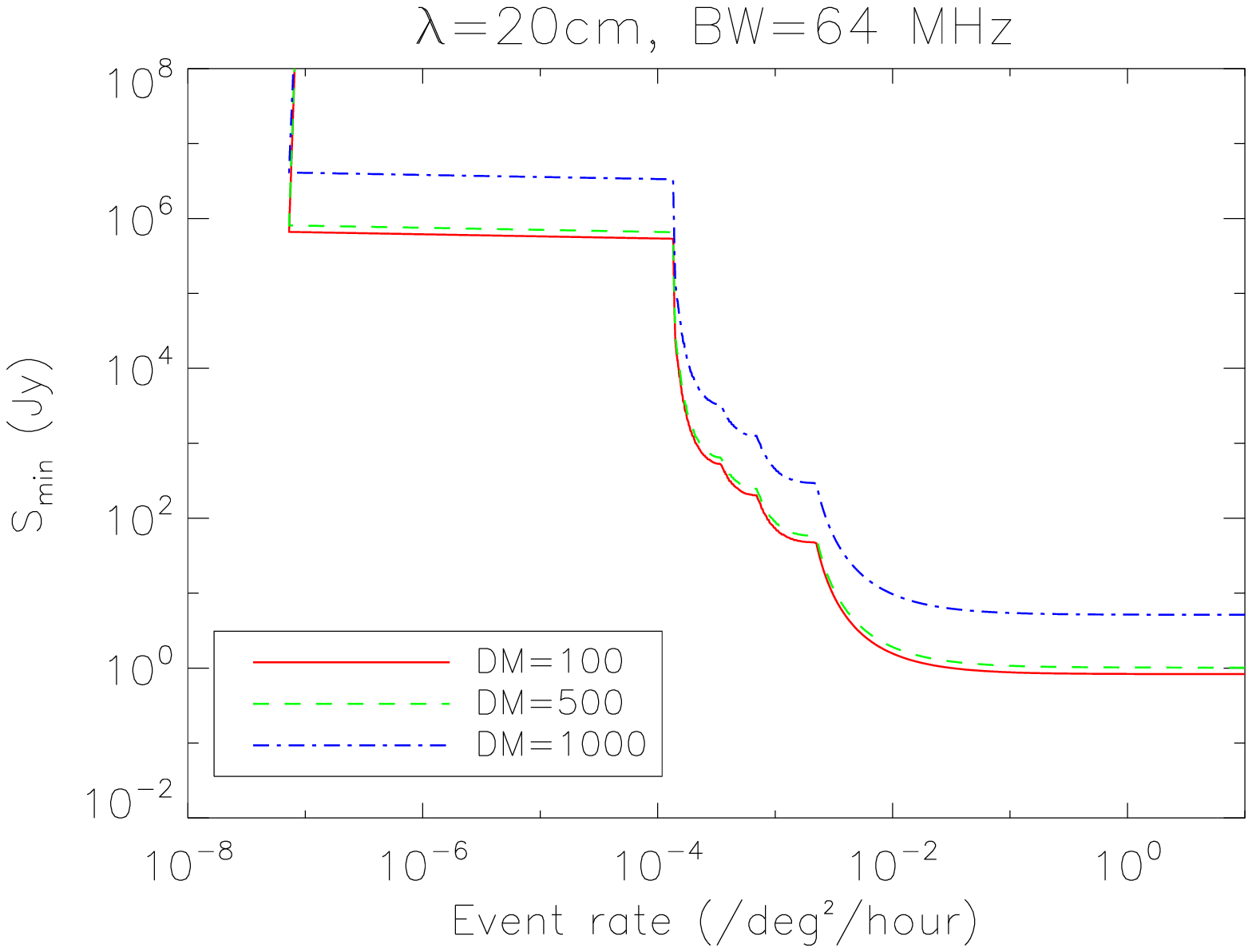}}\\
\subfigure[Variation with bandwidth and observing time (DM = 0), including the previous result from \citet{wayth12} (black, dash-dot-dot-dot).]{\includegraphics[scale=0.75]{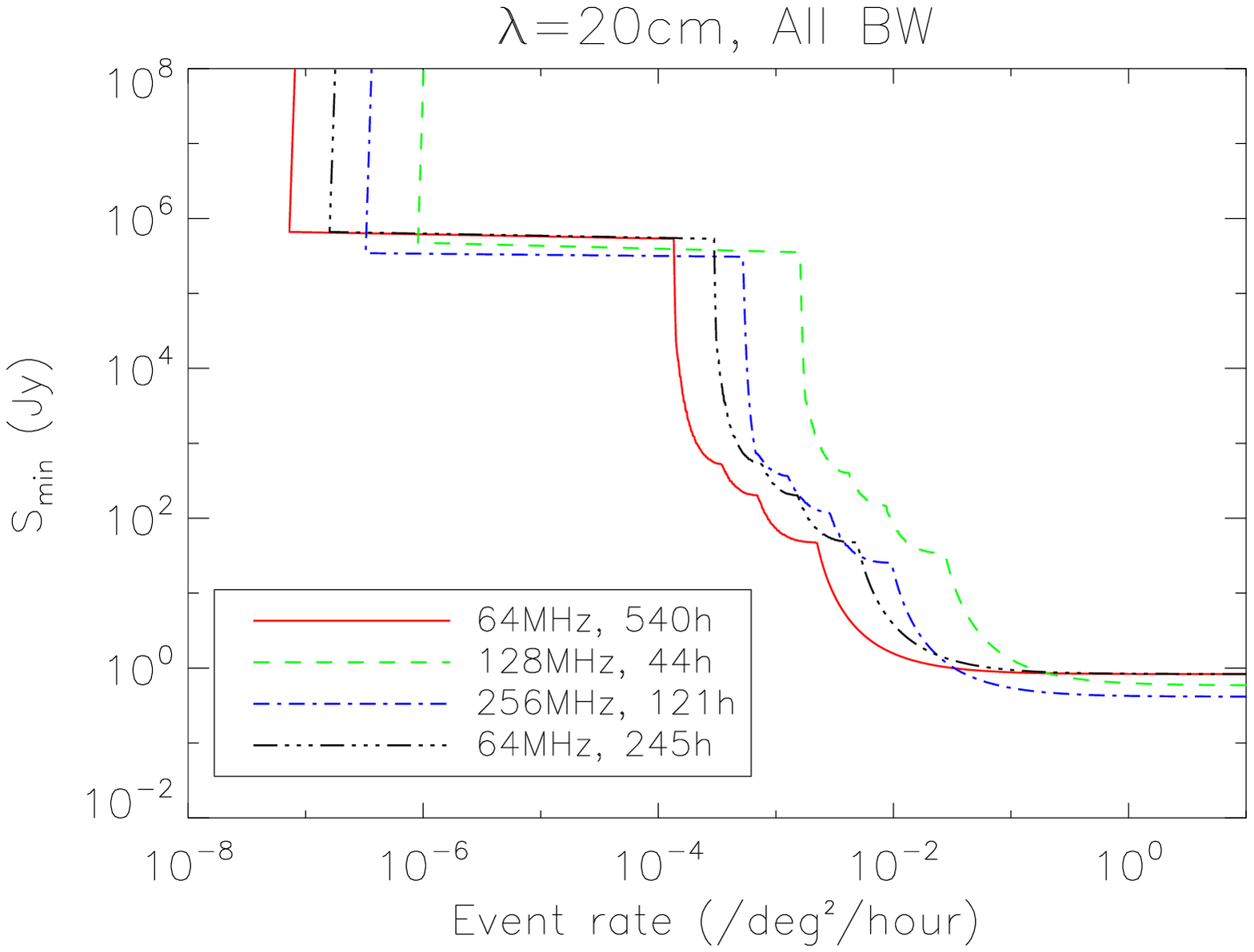}}
\caption{Constraints from recent V-FASTR accumulated data, using the previously-used event rate plot, and the 20~cm receiver. The Airy disk beam has been truncated after the third sidelobe, with constant gain to the horizon beyond that angle. (a) Limits for observations with 64~MHz bandwidth ($\Delta{t}$=540~h) and different source DM. (b) Limits from each bandwidth (DM = 0), and the limit from  \citet{wayth12} (black, dash-dot-dot-dot).}\label{old_event_20cm_fig}
\end{center}
\end{figure}

We first present the latest set of results for the 20~cm receiver using the existing event rate framework, to demonstrate the improvement in time-on-sky compared with the results presented in \citet{wayth12}. Figure \ref{old_event_20cm_fig} displays the constraint curves for $BW$ = 64~MHz and different dispersion measures (a), and for all bandwidths and zero dispersion measure (b). We also plot the existing constraint for $BW$ = 64~MHz.
The curves track as expected --- longer on sky yields stronger constraints on the event rate, while the larger bandwidths yield lower minimum detectable flux densities. The contributions from the three inner sidelobes are evident as cusps in the constraint curves.

\begin{figure}
\begin{center}
\includegraphics[scale=0.9]{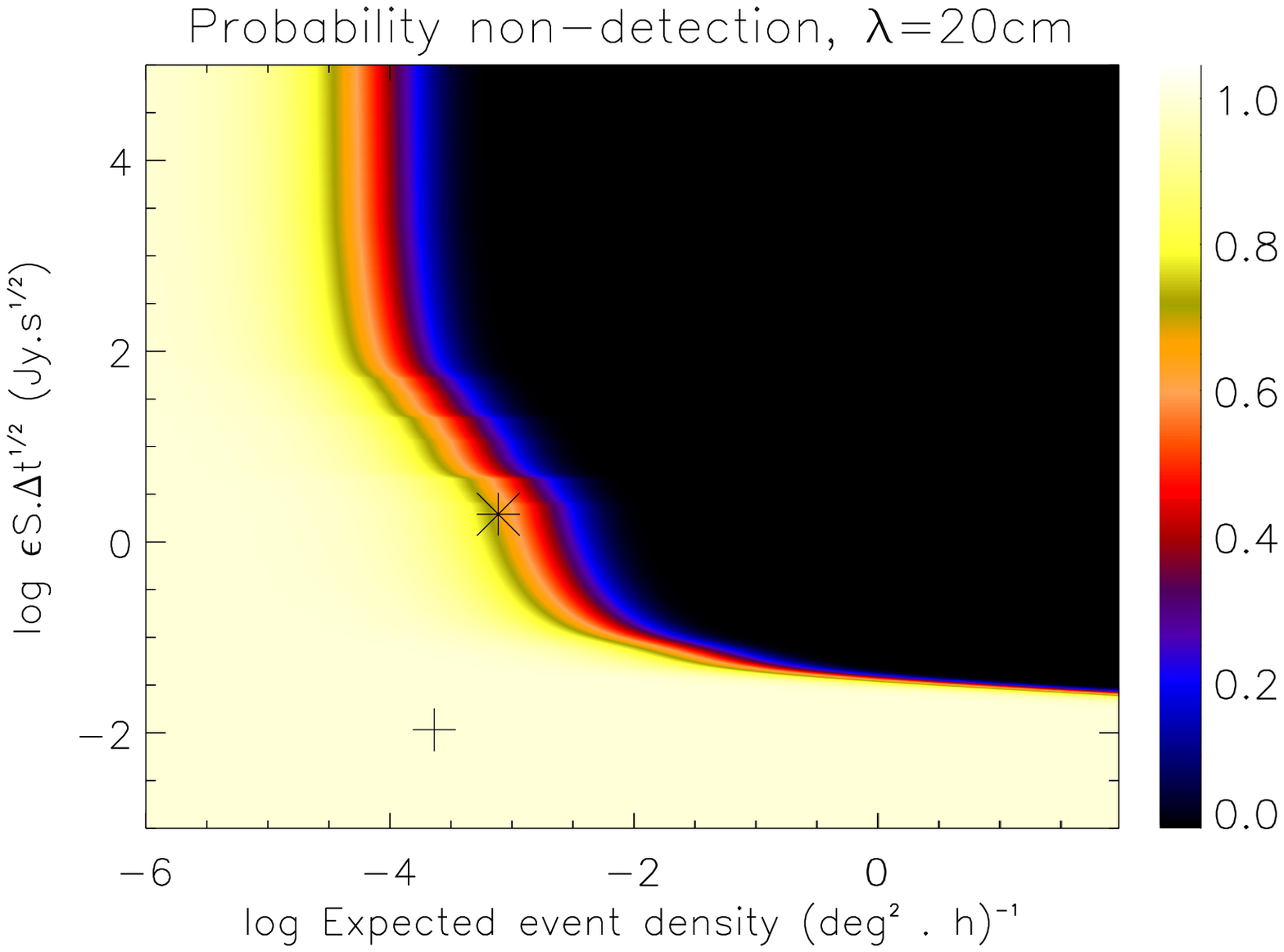}
\caption{Event rate constraint curve for the 20~cm receiver of V-FASTR, using the probability framework and the new detection metric ($\epsilon{S}_{{\rm actual}}\sqrt{W}$). Also plotted are the event rates for the \citet{lorimer07} (asterisk) and \citet{keane11} (plus) experiments, incorporating loss in detection performance due to dispersive temporal smearing.}\label{new_event_20cm_fig}
\end{center}
\end{figure}

Figure \ref{new_event_20cm_fig} displays the probability contours for the 20~cm results from V-FASTR, plotting the probability a candidate in that region of parameter space would \textit{not} be observed by the experiment.
The abscissa is the expected number of events per square degree per hour, and the ordinate is the combined detection quantity: $\epsilon{S}_{{\rm actual}}\sqrt{W}$ (with dimensions, Jy$\sqrt{s}$), formed according to equations \ref{final_equation} and \ref{combined_prob_eqn}. We also display the event rates for two experiments with the Parkes radiotelescope, which both found a candidate fast transient event \citep{lorimer07,keane11,keane10}. The measured DMs for these candidate events have been incorporated into the efficiency factor, $\epsilon$, and used in the plot. The structure of the isocontours is real, and corresponds to the inner sidelobes of the VLBA antennas. Table \ref{lorimer_keane_table} displays the values used to plot these events.
\begin{table}
\begin{center}
\begin{tabular}{|c|cccc|}
\hline Candidate & DM & $\Delta{t}_{s}$ & $S_{\rm peak}$ & $\Delta{t}_{\rm Tot}$.FOV \\ 
\hline Unit &	pc.cm$^{-3}$ &	s &	Jy & deg.$^2$ h\\
\hline \citet{lorimer07} & 375 & 0.005 & 30 & 1303	\\
\citet{keane11} & 745 & 0.007 & 0.41 & 4264\\
\hline 
\end{tabular}
\caption{Event rates from the two candidate events from the \citet{lorimer07} and \citet{keane11} experiments with the Parkes radiotelescope (note that the thirteen multibeam receivers have been incorporated into the survey volume metric, $\Delta{t}_{\rm Tot}$.FOV). We have assumed an Airy disk, and extended to the third null, as with the VLBA antennas.}\label{lorimer_keane_table}
\end{center} 
\end{table}
Interestingly, the event rate implied by the Lorimer event is on the cusp of expected detectability for the 20~cm receiver of V-FASTR, and future data releases will begin to challenge detections in that region of parameter space (assuming no V-FASTR detections).

\begin{figure}
\begin{center}
\subfigure[4~cm receiver.]{\includegraphics[scale=0.45]{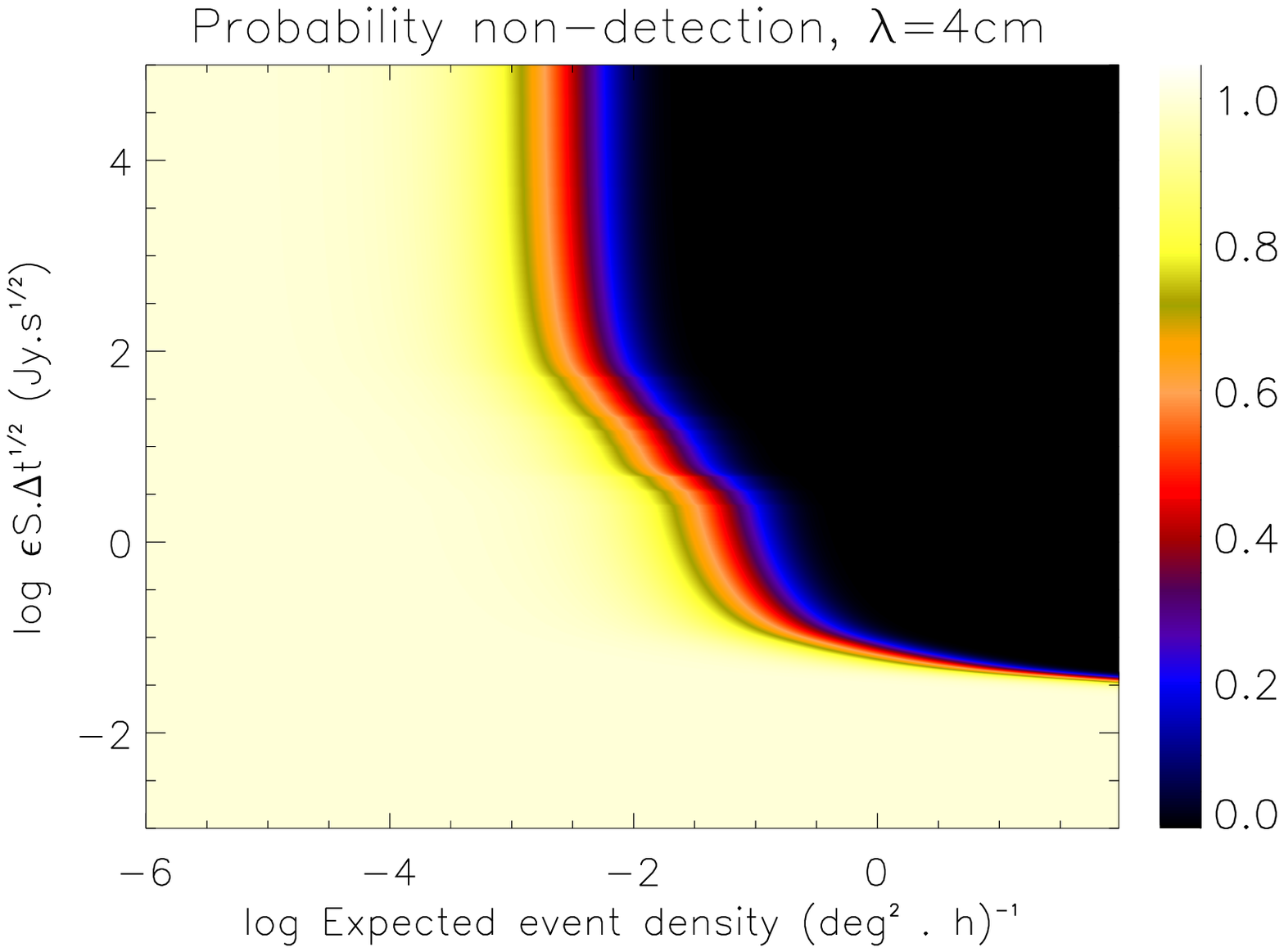}}
\subfigure[2~cm receiver.]{\includegraphics[scale=0.45]{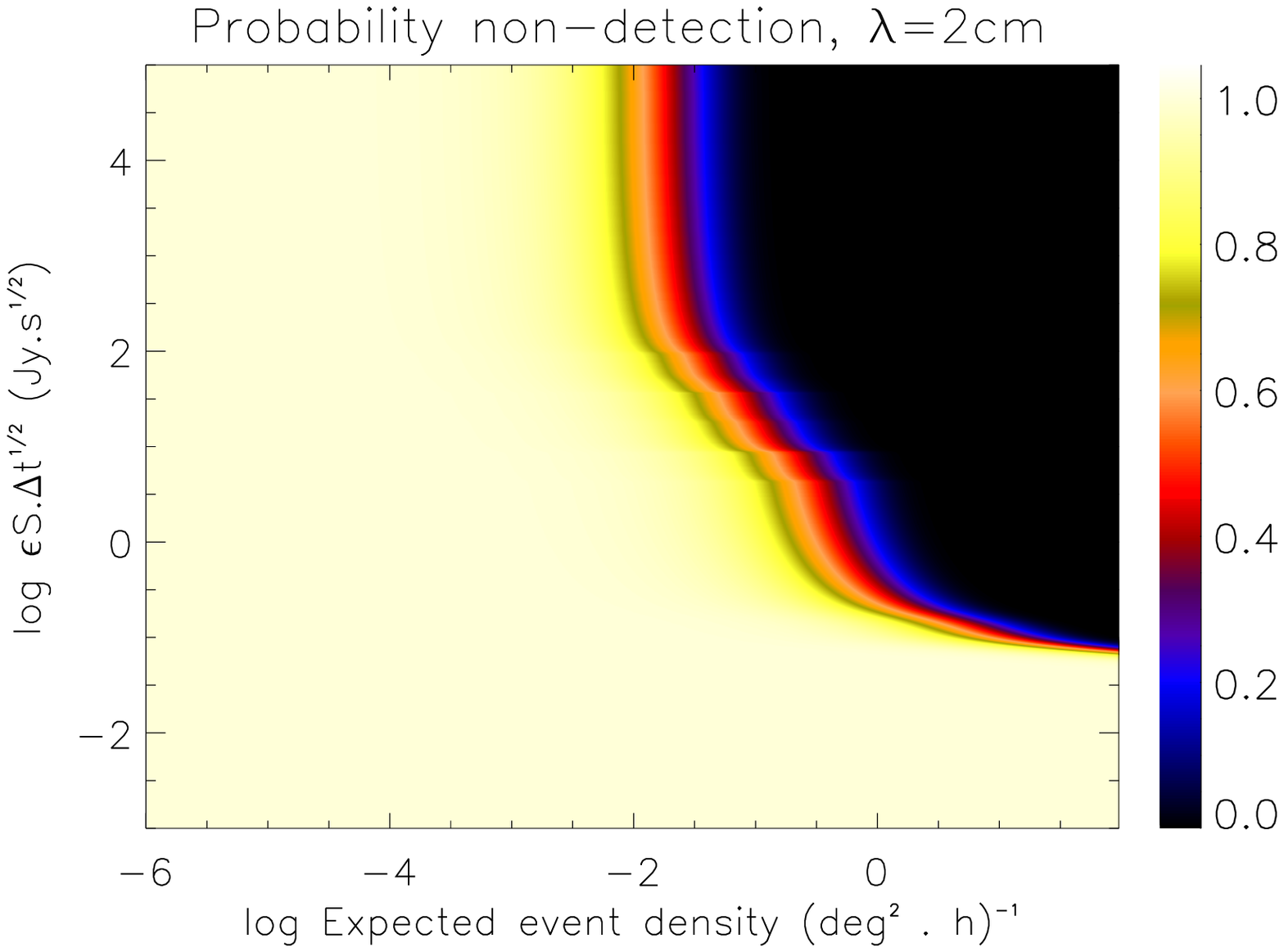}}\\
\subfigure[1~cm receiver.]{\includegraphics[scale=0.45]{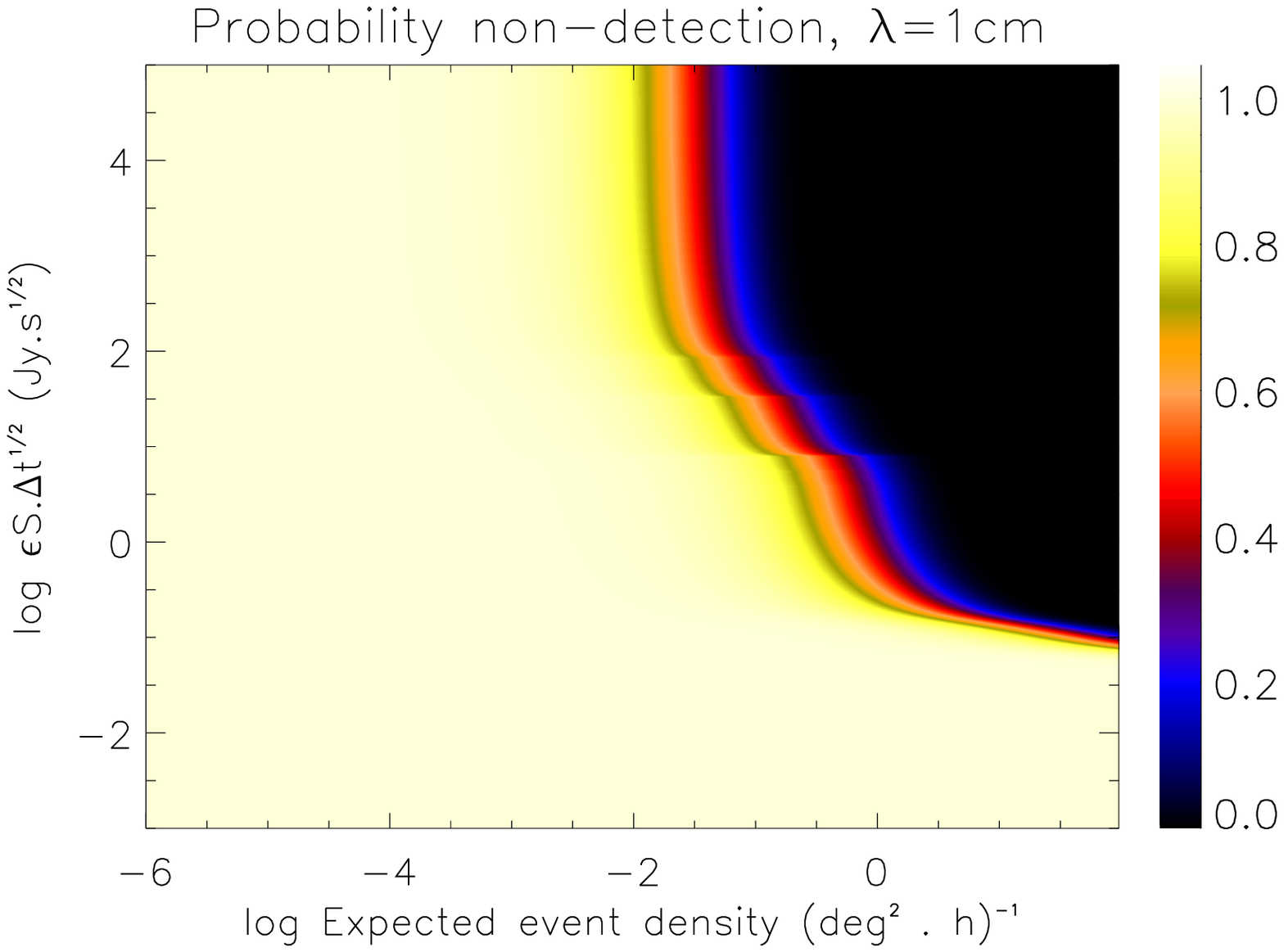}}
\subfigure[7~mm receiver.]{\includegraphics[scale=0.45]{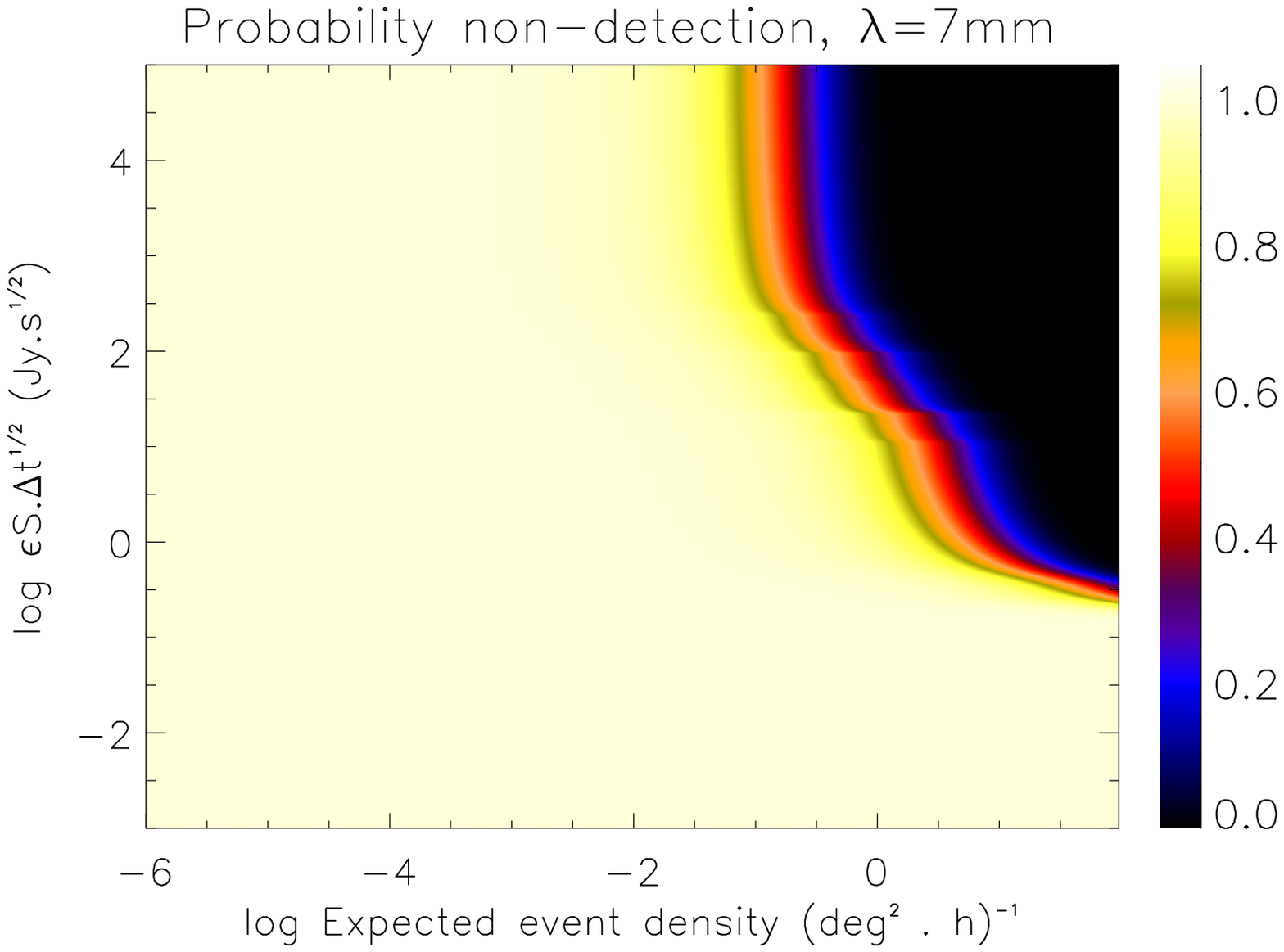}}
\caption{Event rate constraint curves for four interesting wavebands, with the V-FASTR experiment.}\label{new_event_4cm_2cm_1cm_7mm_fig}
\end{center}
\end{figure}

In figure \ref{new_event_4cm_2cm_1cm_7mm_fig} we display the constraints for other interesting V-FASTR receiver bands. All of these wavebands have substantial time-on-sky at the higher bandwidth (256~MHz).
The longer wavelengths sample a larger field-of-view, corresponding to a greater survey volume, and lower constraints for the same time-on-sky. They also have lower system temperatures (SEFD), implying better sensitivity constraints. In particular, the 4~cm receiver has the advantageous combination of long time-on-sky, low system temperature, and large field-of-view, yielding strong constraints at $\sim$8~GHz.

\begin{figure}
\begin{center}
\subfigure[ATA Fly'e Eye constraint curve \citep{siemion12}.]{\includegraphics[scale=0.45]{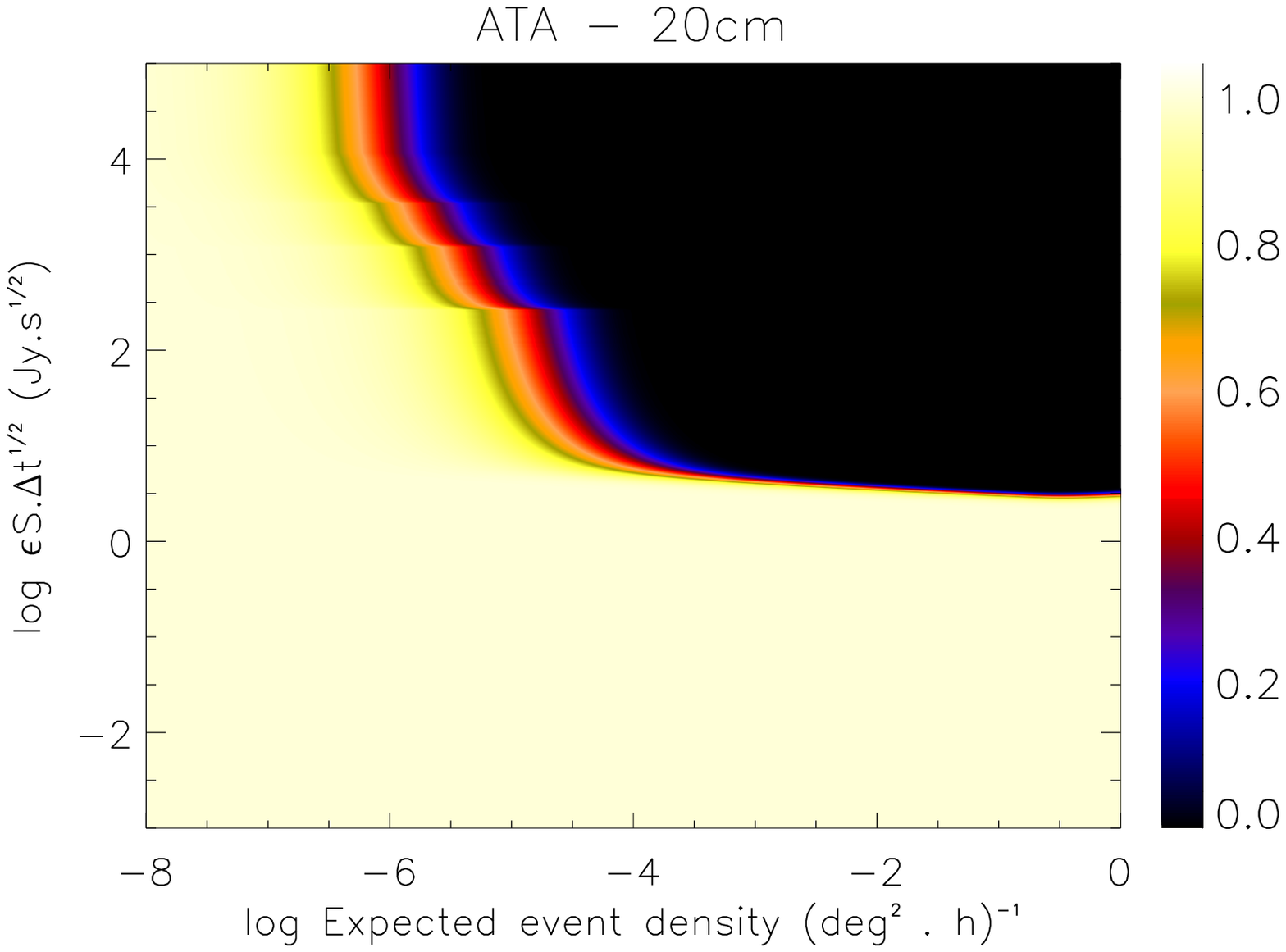}}
\subfigure[Combined 20~cm limit from V-FASTR and ATA.]{\includegraphics[scale=0.45]{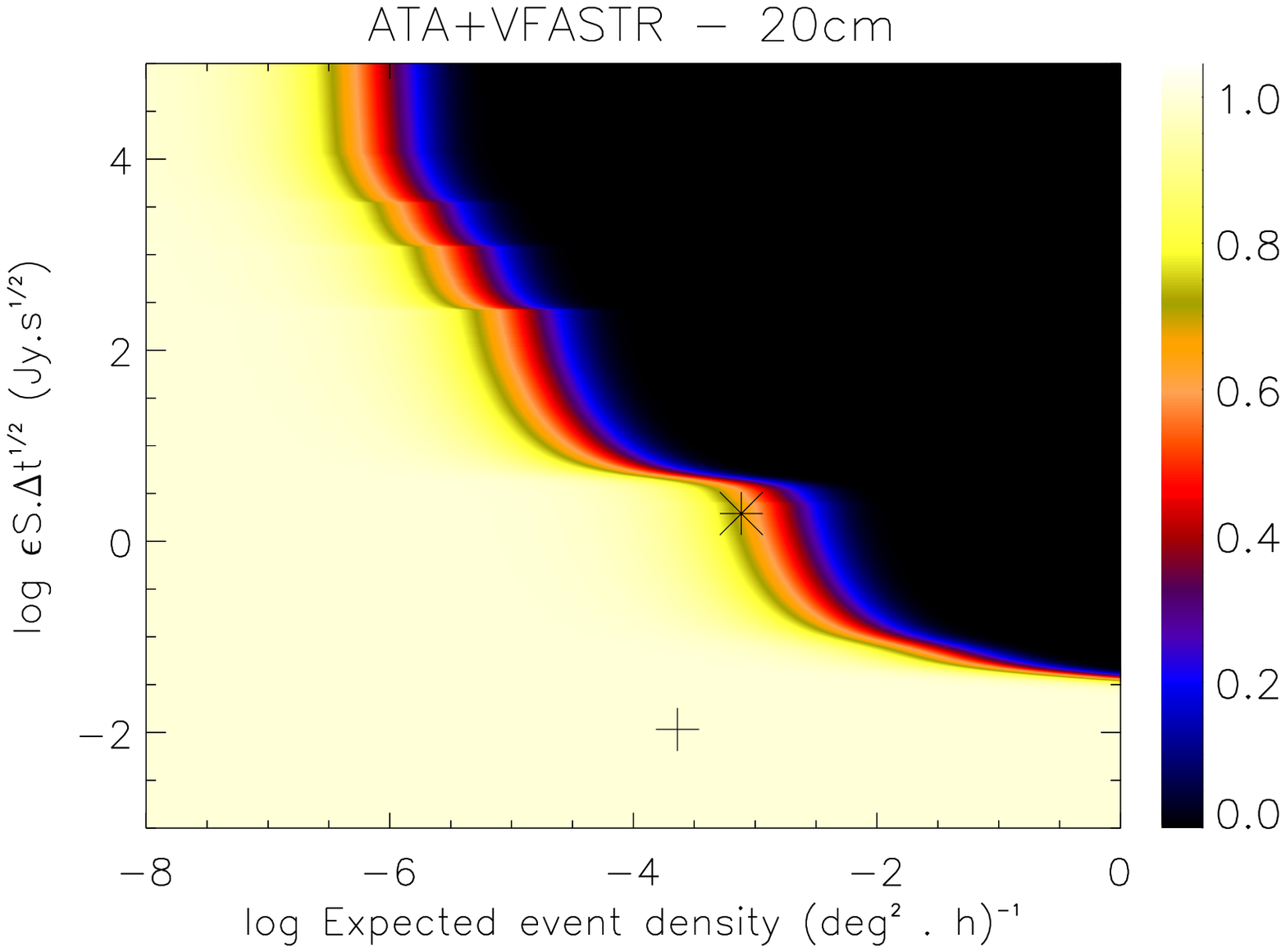}}
\caption{Event rate constraint curves for the ATA Fly's Eye survey, using the new framework \citep{siemion12} (a), and combined with the V-FASTR results (b).}\label{ata_vfastr_fig}
\end{center}
\end{figure}

We now use the new framework to explore the combined constraints from different surveys. As discussed previously, it is not reasonable to form a sensible comparison of single-element and multiple-element experiments, due to the different RFI mitigation strategies one must employ. Instead, we choose to combine the V-FASTR results with those from the ATA Fly's Eye experiment \citep{siemion12}. In `Fly's Eye' mode, the ATA employs 30 of its 6-m dishes, each pointed in a different direction, leading to large instantaneous fields-of-view, but limited sensitivity\footnote{Assuming an Airy disk for the ATA beam shape, which may not be a good approximation. Different beam shapes will alter the details of the cusps in the event rate diagram, but not the overall shape (which is set by the dish size).}. This is in constrast to V-FASTR, where the incoherent combination of dishes lead to good sensitivity, but limited field. Figure \ref{ata_vfastr_fig} displays the event rate constraint diagrams for the ATA alone (a), and for the combined ATA$+$V-FASTR experiments.
The different strengths of each experiment combine to form a strong constraint across much of parameter space. Again, we also plot the results from \citet{lorimer07} and \citet{keane11}. The limited sensitivity of the ATA dataset cannot add significant information at these flux density levels.

\subsection{Fast transients constraints with SKA Phase I}
\begin{figure}[ht]
\begin{center}
{\includegraphics[scale=0.8]{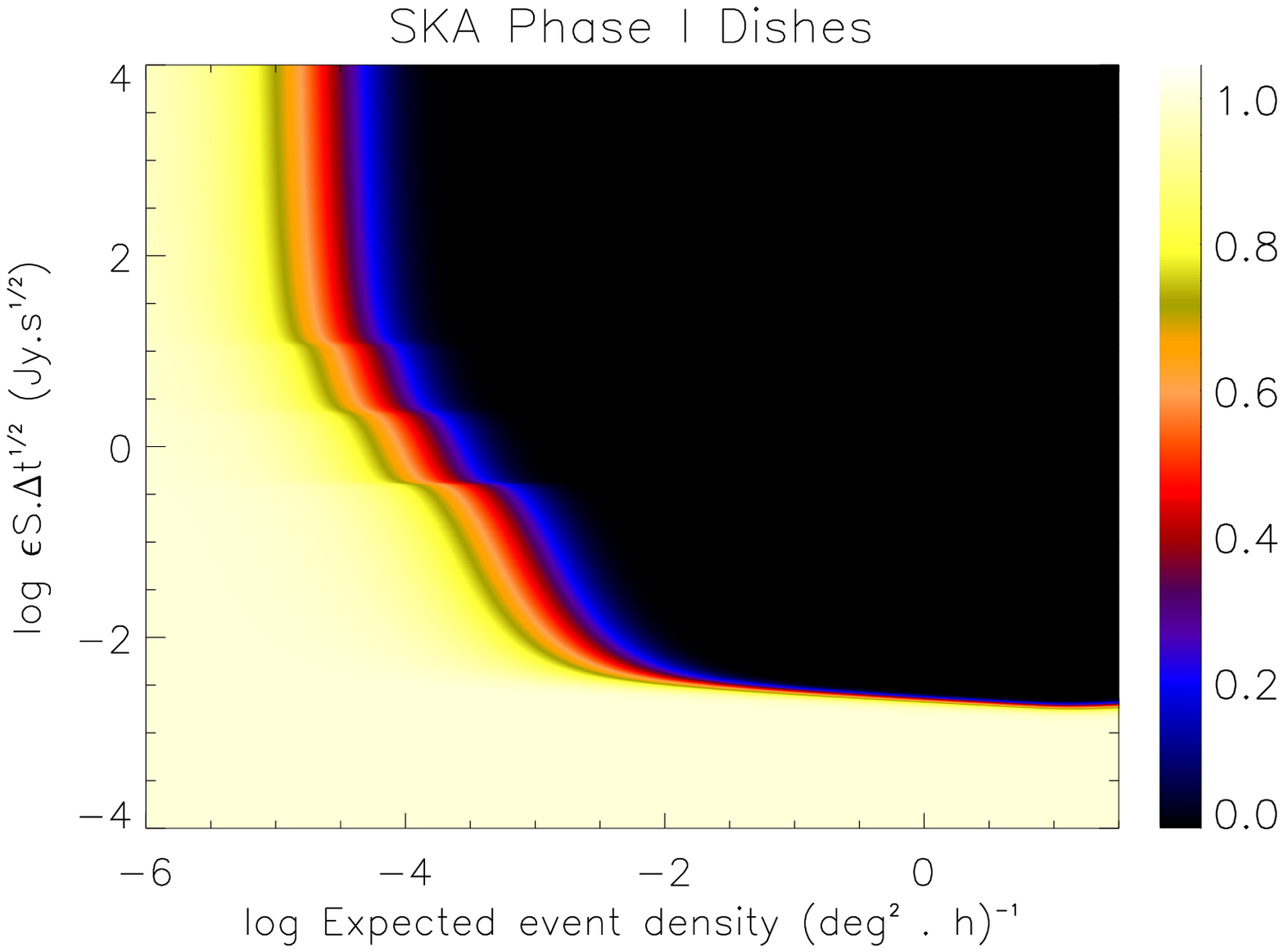}}
\caption{Event rate constraint curves for the SKA Phase I dish array, assuming 1~GHz bandwidth between 1 and 2~GHz, 250 dishes in incoherent mode, and 461 hours of observation \citep[following system specifications of][]{dewdney10}.}\label{ska_fig}
\end{center}
\end{figure}

The system specifications for the SKA Phase I dish array outline an instrument with 250 dishes distributed over an area with a maximum baseline of 100~km \citep{dewdney10}. In the high-band (1-2~GHz), the description specifies a maximum bandwidth of 1000~MHz, yielding an instrument with substantially different beam shape characteristics at each end of the band. Following the work in \citet{wayth12}, we generate an event rate constraint curve for the system specifications in \citet{dewdney10} in the high-band, using an incoherent addition of antennas, and 461 hours of observation (figure \ref{ska_fig}).
The sidelobe pattern is again observed in the structure of the isocontours. The plot deviates from that shown in \citet{wayth12}, due mostly to the significant fractional bandwidth, and the correct accounting for sensitivity variation with frequency and angle from the boresight. In addition, the SKA Phase I dish dump rate is 0.1~s (due to limitations imposed by the long baselines), yielding degraded performance for fast transients on shorter timescales. This is incorporated into the metric plotted here. Compared with the ATA, the SKA has substantially better sensitivity, but poorer field-of-view in incoherent addition mode. SKA dishes may also be deployed in other fast transient configurations, and these will have different limits \citep{colegate12}.

\section{Conclusions}
We present the latest set of results from the V-FASTR experiment, a commensal fast transient experiment using the VLBA network across ten wavebands. These results are presented through the lens of a new framework, developed to incorporate source and experimental parameters into a metric that (1) captures additional information that is meaningful for characterizing the true sensitivity of an experiment, and (2) allows independent datasets to be combined in a rigorous manner. We use this framework to combine results from different bandwidth observations within V-FASTR, demonstrating that the 20~cm receiver observations are pushing into the parameter space defined by the Lorimer burst and Keane event \citep{lorimer07, keane11}. We then combine V-FASTR results at 20~cm with the ATA Fly's Eye dataset, demonstrating the complementary nature of the two experiments. Finally we present expectations for the SKA Phase I dish array instrument, producing plots that deviate from those presented previously, due to the correct accounting for beam shapes as a function of frequency and angle from the boresight.

\acknowledgments{The Centre for All-sky Astrophysics is an Australian Research Council Centre of Excellence, funded by grant CE110001020. The International Centre for Radio Astronomy Research (ICRAR) is a Joint Venture between Curtin University and the University of Western Australia, funded by the State Government of Western Australia and the Joint Venture partners. SJT is a Western
Australian Premiers Research Fellow. RBW is supported via the Western Australian Centre of Excellence in Radio Astronomy Science and Engineering. ATD was supported by
an NRAO Jansky Fellowship and an NWO Veni Fellowship. Part
of this research was carried out at the Jet Propulsion Laboratory,
California Institute of Technology, under contract with the US
National Aeronautics and Space Administration. The National
Radio Astronomy Observatory is a facility of the National
Science Foundation operated under cooperative agreement by
Associated Universities, Inc. This research has made use of
NASA’s Astrophysics Data System.\\
Facility: VLBA
}



\newpage
\bibliographystyle{jphysicsB}
\bibliography{pubs.bib}

\end{document}